\documentclass[a4paper,twoside]{article}
\usepackage{amsmath, amssymb, amsthm, amstext}
\usepackage{epsfig}
\usepackage{caption}
\usepackage{subcaption}
\usepackage{multirow}

\usepackage{calc}
\usepackage{multicol}
\usepackage{pslatex}
\usepackage{apalike}
\usepackage{listings}
\usepackage{cite}
\usepackage{dsfont}
\usepackage{pgfplots}
\pgfplotsset{compat=1.16}
\usepackage{xcolor, soul}
\usepackage{url}

\usepackage[inline]{enumitem}
\usepackage{import}
\usepackage{adjustbox}
\usepackage{tikz}
\usepackage{colortbl}
\usepackage{appendix}

\pgfmathdeclarefunction{gauss}{2}{%
  \pgfmathparse{1/(#2*sqrt(2*pi))*exp(-((x-#1)^2)/(2*#2^2))}%
}
\newcommand\gauss[2]{1/(#2*sqrt(2*pi))*exp(-((x-#1)^2)/(2*#2^2))}

\def\checkmark{\tikz\fill[scale=0.2](0,.35) -- (.25,0) -- (1,.7) -- (.25,.15) -- cycle;} 

\usetikzlibrary{backgrounds,calc,shadings,shapes,arrows,arrows.meta,shapes.arrows,shapes.symbols,patterns, datavisualization, datavisualization.formats.functions, matrix}
\usepackage[vlined,ruled]{algorithm2e}

\usepackage{SCITEPRESS} % Please add other packages that you may need BEFORE the SCITEPRESS.sty package.

\begin{document}

% --------- Title  ------------------------
\title{Dynamic Link Network Emulation: a Model-based Design}

\author{\authorname{Erick Petersen\sup{1,2}, Jorge L\'opez\sup{1}, Natalia Kushik\sup{2}, Claude Poletti\sup{1} and Djamal Zeghlache\sup{2}}
\affiliation{\sup{1}Airbus Defence and Space, Issy-Les-Moulineaux, France}
\affiliation{\sup{2}T\'el\'ecom SudParis, Institut Polytechnique de Paris, Palaiseau, France}
\email{\{erick.petersen, jorge.lopez-c, claude.poletti\}@airbus.com, \\
\{erick\_petersen, natalia.kushik, djamal.zeghlache\}@telecom-sudparis.eu}
}

% --------- Keywords ------------------------


% ---------Abstract-------------------

\abstract{
This paper presents the design and architecture of a network emulator whose links' parameters (such as delay and bandwidth) vary at different time instances. The emulator can thus be used in order to test and evaluate novel solutions for such networks, before their final deployment. To achieve this goal, different existing technologies are carefully combined to emulate link dynamicity, automatic traffic generation, and overall network device emulation. The emulator takes as an input a formal model of the network to emulate and configures all required software to execute live software instances of the desired network components, in the requested topology. We devote our study to the so-called \emph{dynamic link networks}, with potentially asymmetric links. Since emulating asymmetric dynamic links is far from trivial (even with the existing state-of-the-art tools), we provide a detailed design architecture that allows this. As a case study, a satellite network emulation is presented. Experimental results show the precision of our dynamic assignments and the overall flexibility of the proposed solution.}

\onecolumn \maketitle \normalsize \setcounter{footnote}{0} \vfill

% --------- Sections  -----------------

\section{\uppercase{Introduction}}
\label{sec:introduction}

As the demand for interactive services, multimedia and network capabilities grows in modern networks, novel software and/or hardware components should be incorporated \cite{deng2019next}. As a consequence, the evaluation and validation process of these newly developed solutions is critical to determine whether they perform well, are reliable and robust before their final deployment in a real network \cite{9275875}. However, thorough testing or qualifying \cite{shan2021testing} the produced software under a wide variety of network characteristics and conditions is a challenging task \cite{gandhi2021design}.

Currently, many of these tests are done through operational, controlled and small-scale networks (physical testbeds) or alternatively software-based testbeds. Ideally, if available, such tests are performed on the original system in order to replicate the conditions in which a service or protocol will be used at the highest level of fidelity \cite{6805124}. Unfortunately, while system modeling is not needed, such testbeds are not always desirable or pertinent due to several reasons \cite{sun2003satellite}. 
% Below, we list some of those:
%\begin{itemize}
%    \item difficulty in creating the test environment due to the unavailability or high costs of the necessary equipment for them (for example, a geostationary satellite, if available, could be costly);
%    \item difficulty in controlling the test environment under specific conditions; when dealing with real equipment, there are elements that are beyond the control of the researchers (such as queue sizes, various delays, link quality due to network load or weather conditions that may affect the radio-physical links in specific network technologies);
%    \item low flexibility since a small modification in the test environment (e.g., a small network topology or link property change) often translates to a physical change which implies additional time and costs.
%\end{itemize}
For example, there are difficulties in creating various network topologies, generating different traffic scenarios, and testing the implementations under specific conditions (network load or weather conditions that may affect the radio-physical links in specific network technologies such as wireless or satellite communications).

A very well-known alternative method is the use of network simulators \cite{6487111}. Through network simulation, researchers can mimic the basic functions of network devices and study specific network-related issues on a single computer or high-end server. However, the adequacy of simulated systems is always in question due to the model abstraction and simplification. % Furthermore, the behavior of the simulated components is restricted only to static changes and to the models employed for each simulated characteristic. This could result in a loss of accuracy if any non-deterministic component (within the scope of modelling) is included in the simulation. For example, the essence or the main idea behind a novel routing algorithm for a satellite communication network can generally be validated using well established simulators such as Opnet or NS-3. But once, it gets implemented as a final service or application such simulators cannot be used anymore to test it. 
At the same time, not simulation but emulation for the related networks can also be a solution \cite{10.1007/978-3-030-32216-8_49}. Network emulation provides the necessary mechanisms to reproduce the behavior of real networks at low infrastructure costs, compared with physical testbeds while achieving better realism than simulations since it allows the interaction with interfaces, protocol stacks and operating systems. Moreover, there is a possibility to perform continuous testing on the final implementation without having to make any changes in the solution once deploying it in a real network. % However, although network emulation offers a great balance between scalability and fidelity, it is still a major concern to verify the adequacy of such emulators and its components for a given context. An uncontrollable and poorly specified environment in this case could lead to slow experimental cycles, non-reproducible results, and potential downstream costs in future implementations. In addition, we argue that 
However, the emulation of dynamic link networks, i.e., networks whose link parameters change, complicates the emulation architecture. For example, certain radio-frequency links have different up/down bandwidth capacity \cite{gandhi2021design}, large delays (due to distant transmitters), and the links' capacities may change due to external interference, propagation conditions (weather), traffic variations (due to the shared medium), or others.  Therefore, it is extremely important to have methods which allow controlling key parts of the emulation over time, such as the generation of traffic or the modification of the link property values (capacity, delay). These are required in order to build a proper emulation environment of interest which is the main focus of this work. 

To cope with such requirements, we herein propose a dynamic link network emulation and traffic generation which combines the functional realism and scalability of virtualization and link emulation to create virtual networks that are fast, customizable and portable. The main contributions of this paper are: \textbf{i)} a design and architecture of a dynamic link network emulator that meets current and future link emulation needs by the effective use of software technologies, such as virtualization (containers and virtual machines) and Linux kernel capabilities; \textbf{ii)} a dynamic link network emulator that provides a fast and user friendly workflow, from the installation to the configuration of scenarios by using a formal model of the network; % in terms of first order logic formulas and verified throughout a satisfiability modulo theories solver providing the accuracy and flexibility of a real environment.
     % update according the evaluation or experiments
 \textbf{iii)} a use case illustrating the dynamic link capabilities of our emulator; we provide an experimental evaluation, varying different network parameters configured by demand for effectively emulating satellite communications.  
    %\item A dynamic link network emulator that is generic enough (applied to any case study, including satellite communications), portable and scalable from a single computer to a cloud system to qualify any application or transport level implementations. 

 %We note that, having a formal model for the network emulation can also serve for verifying that the emulation behaves as the original system with respect to a set of (physical) properties. However, such verification is out of the scope of this paper; the interested reader may refer, for example to \cite{emul_verif_nca2020} where the corresponding issues have been raised.

\section{\uppercase{Related Work}}\label{sec:relwork}
Several works have been devoted to simulation and emulation of networks,  to perform experiments on novel or existing protocols and algorithms. Below, we briefly summarize some existing solutions.

Ns-3 \cite{ns-3} is a widely used network simulator. Ns-3 simulates network devices by compiling and linking C++ modules; thus it simulates the behavior of real components in a user-level executable program. However, real world network devices are extremely complex (functionally speaking) or cannot be compiled and linked together with Ns-3 to form a single executable program. Therefore, Ns-3 cannot run real-world network devices but only specific ones developed for it. vEmulab \cite{stoller2008large} is a network emulator with a minimum degree of virtualization, aiming to provide application transparency and to exploit the hierarchy found in real computer networks. Its architecture uses FreeBSD jail namespaces \cite{kamp2000jails} in low-end computers to emulate virtual topologies. Similarly, Mininet \cite{lantz2010network} enables rapid testbeds by using several virtualization features, such as virtual ethernet pairs and processes in Linux container network namespaces. It emulates hosts, switches and controllers which are simple shell processes that are given their network namespace and links between them. However, both still present some limitations including the lack of support for dynamic features such as link emulation, resource management and traffic generation. %In contrast, virtualization (both  hypervisor-based  and container-base) allows our platform to provide resource isolation and limitation, support for dynamic services, applications and protocols through pre-built images. Moreover, it permits to run different operating systems adding portability. 
EstiNet \cite{wang2013estinet} is based on network simulation/emulation integration for different kinds of networks. Unlike previous simulators, EstiNet allows not only observation but also configuration through a GUI. It also supports wireless channel modeling. However, since EstiNet is a commercial solution, it cannot be easily extended. Moreover, its features are limited and depend on the EstiNet developers. Thus, the performance fidelity and the expansion to new features are reduced. OpenNet \cite{chan2014opennet} also merges simulation and emulation network capabilities by connecting Mininet and ns-3. However, it inherits the limitations from Ns-3 and Mininet. Additionally, its main focus is on software-defined wireless local area networks (SDWLAN). 

\begin{table*}[!ht]
\centering
\begin{center}

% A table with adjusted row and column spacings
% \setlength sets the horizontal (column) spacing
% \arraystretch sets the vertical (row) spacing
\begingroup
\setlength{\tabcolsep}{8pt} % Default value: 6pt
\renewcommand{\arraystretch}{1} % Default value: 1
\setlength\doublerulesep{.1pt}

%\small
%\footnotesize
\scriptsize
\begin{tabular}{
| 
>{\raggedright\arraybackslash}p{2.5cm} |
>{\raggedright\arraybackslash}p{0.6cm}|
>{\raggedright\arraybackslash}p{1cm}|
>{\raggedright\arraybackslash}p{0.3cm}  |
>{\raggedright\arraybackslash}p{0.9cm}  |
>{\raggedright\arraybackslash}p{0.9cm}  |
>{\raggedright\arraybackslash}p{0.9cm}  |
>{\raggedright\arraybackslash}p{0.7cm}  |
>{\raggedright\arraybackslash}p{1.1cm}  |
>{\raggedright\arraybackslash}p{1.2cm}  |
} 

\hline

% license type  ( opensource, comercial )
% language ( c++, python, java ; OTCL)
% GUI  ( yes, no, limited)
% Emulation support (no; yes; limited support)
% Scalability ( small, medium, large)
% Portability ( small, medium, large)

% Dynamic Link Support
% Real Traffic Generation Support
% Formal Description (verification)

% \hline
\rowcolor{gray}
 Name & 
 Open Source &  %Open source
 Language & 
 GUI & 
 Emulation Support & 
 Scalability & 
 Portability  & 
 Dynamic Link & 
 Automatic Traffic Generation & 
 Formal Description
\\
%& \begin{tabular}{@{}c@{}}Form \\ Descr \end{tabular} \\

 \hline
 Ns-3\cite{ns-3} & \checkmark & C++/Python & x & x & +++ & x & \checkmark & x & x\\
 %\hline
 %Ns-4\cite{fan2017ns4}	& \checkmark  & C++/Python & x & x & +++ & x & \checkmark & x & x\\
 \hline
 Mininet\cite{mininet} & \checkmark  & Python & \checkmark & \checkmark & + & \checkmark & x & x & x\\
 \hline
 Containernet\cite{containernet} & \checkmark  & Python & \checkmark & \checkmark & ++ & \checkmark & x & x & x\\
 \hline 
 OMNet++\cite{omnet} & x & C++ & \checkmark & x & +++ & x & \checkmark & x & x\\ 
 \hline
 Emulab\cite{stoller2008large}& \checkmark & C & \checkmark &  \checkmark & + & \checkmark & x & x & x\\
 \hline
 OpenNet\cite{chan2014opennet} & \checkmark & C++ & \checkmark & x & +++ & x & \checkmark & x & x \\
 \hline
 vSDNEmul\cite{farias2019vsdnemul}& \checkmark & Python & x & \checkmark & ++ & \checkmark & x &x  & x\\
 \hline
 EstiNet\cite{wang2013estinet}& x & - & \checkmark & \checkmark & ++ & \checkmark & \checkmark & - & -\\
 \hline
 SDN Owl\cite{srisawai2018rapid} & \checkmark & Python & x & \checkmark & ++ &\checkmark & x & x &x\\
% \hline
% Mesh Linux Containers (MLC)\cite{6379145} & \checkmark & -  & x & \checkmark & x & x & x & x & x\\
% \hline
% CORE (Common open research emulator)\cite{ahrenholz2008core} & \checkmark & Python & \checkmark & \checkmark & + & \checkmark & \checkmark & x& x\\ 
% \hline
%Cloonix3\cite{neumann2008better}& \checkmark & C & \checkmark & \checkmark & + & x & x & x & x\\
% \hline
%  GNS3\cite{neumann2015book} & \checkmark & Python & \checkmark & \checkmark & - & x & x & x & x\\ 
 \hline
 NetEM\cite{hemminger2005network}& \checkmark & - & x & \checkmark & - & x & \checkmark & x & x\\
% \hline
%  WANem\cite{kalitay2011designing}  & \checkmark & - & \checkmark& \checkmark & + & \checkmark & x & x & x\\
% \hline
% NEMAN\cite{puvzar2005neman} & \checkmark & Python & \checkmark & \checkmark & ++ & \checkmark & \checkmark & x & x\\
% \hline
 %MiniWorld\cite{schmidt2017miniworld} & \checkmark & Python & x & \checkmark & + & \checkmark & x &  x & x\\
 \hline
 
% \hline
% Ns-1\cite{mccannens}  & 3\checkmark & + & - & +/-/\checkmark & \\ [0.5ex] 
% \hline
% Ns-2\cite{issariyakul2009introduction} & 545 & 778 & 7507 &  33 & 22\\
 % \hline
% BAMNE\cite{britos2015batman}   & \checkmark & + & - & +/-/\checkmark & \\
%  \hline
%  Netkit\cite{pizzonia2008netkit} 	  & \checkmark & + & - & +/-/\checkmark & \\
 %\hline
 %MobiREAL\cite{1521172}  & \checkmark & + & - & +/-/\checkmark & \\
% \hline
% JiST/SWANs\cite{Barr:2005:JEA:1060168.1060170} & \checkmark & + & - & +/-/\checkmark & \\
% \hline
% S3F/S3FNet\cite{jin2014parallel}& \checkmark & + & - & +/-/\checkmark & \\

\end{tabular}
\endgroup

\end{center}
\caption{Comparison of Software-based Network Testbeds}
%\caption{Comparison of Software-based Network simulators/emulators}
\label{tab:survey}
\end{table*}

More recently, the introduction of lightweight virtualization technologies (e.g., containers) has led to some few container-based emulation tools \cite{srisawai2018rapid, farias2019vsdnemul}. SDN Owl \cite{srisawai2018rapid} is a network emulation tool to create simple SDN testbeds using few computers with Linux OSs. SDN Owl utilizes Ansible to send a set of scripts to properly configure each virtual component. However, it fails to provide scalability and isolation, since experiments with different types of network topologies or resource allocation are not shown. vSDNEmul \cite{farias2019vsdnemul} and ContainerNet\cite{containernet} are network emulators based on Docker container virtualization allowing autonomous and flexible creation of independent network elements, resulting in more realistic emulations. However, these emulators can only create SDN networks. Furthermore, the network descriptions remain  rather informal and do not facilitate the verification of the emulator, in order to guarantee that the emulator properly replicates the desired network. A feature comparison is shown in Table~\ref{tab:survey}.

As can be seen and to the best of our knowledge, we are not aware of any works that meet all the features required to properly emulate dynamic link networks in order to qualify novel engineered solutions. This is the main motivation behind this work, which presents the design and architecture of an emulator that meets all the requirements. In order to do so, a model-based engineering approach is well suited, and thus, utilized.

\iffalse

We categorize network simulators and emulators \cite{bakare2019review} along several dimensions as shown in Table~\ref{tab:survey}.

some parameterse to compare Topo flexibility, link realism, traffic realism, reource realism , OS realism, Functional realism, Easy replication, scalability.

Topology flexibility: The platform should be able to easily create experiments with different topologies or even dynamically changing topologies at runtime.

Traffic realism:The platform should be able to generate and receive real, interactive network traffic to and from the Internet/local network. 

The traffic between two hosts should go through network devices (switches or routers) the same way as in the real world.

Link realism: The platform should be capable of controlling the link quality of each link, such as delay, bandwidth, drop rate, etc., according to the real world link quality trace.

Resource realism:The platform should be able to emulate heterogeneous devices by allocating isolated computing resources to different hosts based on their actually available resources.

OS realism:The platform should be able to emulate devices with their real OS. 
%maybe later 

\fi

\section{\uppercase{Background}}\label{sec:bg}
\paragraph{Dynamic Link Networks}

  As introduced in \cite{emul_verif_nca2020}, a static network is a computer network where each link has a set of parameters that do not change, for example bandwidth (capacity) or delay. Differently from static networks, the parameters of the links may change in dynamic link networks (in the scope of our current work, we assume the network topology does not change); such change can be the consequence of the physical medium (e.g., in wireless / radio frequency networks) or due to logical changes (e.g., rate limiting the capacity of a given link). 
 
\begin{figure}[!hbt]
    \centering
    \begin{tikzpicture}[node distance=5.5cm,>=stealth',bend angle=45,auto,scale=.8, every node/.style={scale=.8}]
    \tikzstyle{switch}=[circle,thick,draw=black!75, inner sep=0.075cm]
    \tikzstyle{host}=[circle,thick,draw=black!75,fill=black!80,text=white, inner sep=0.075cm]
    \tikzstyle{undirected}=[thick]
    \tikzstyle{directed}=[thick,->]

    \node[switch] (s1) {$1$};
    \node[switch, right of=s1] (s2) {$2$};
    \node[switch, below of=s1] (s3) {$3$};
    \node[switch, below of=s2] (s4) {$4$};
   
    \path   
            (s1)    edge[directed, bend left=10]    node[above] {\scriptsize $b(e)=\{4,5,6\}, d(e)=\{1,2\}$}   (s2)
                    edge[directed, bend right=10]    node[above, rotate=90] {\scriptsize $b(e)=\{2,3,4\}, d(e)=\{9,10\}$}   (s3)
                    edge[directed, bend left=10]    node[above, rotate=-45] {\scriptsize $b(e)=\{2,3,4\}, d(e)=\{9,10\}$}   (s4)
            (s2)    edge[directed, bend left=10]    node[below] {\scriptsize $b(e)=\{2,3,4\}, d(e)=\{9,10\}$}  (s1)
                    %edge[directed, bend right=10]    node[above, rotate=45] {$b(e)=\{2,3,4\}, d(e)=\{9,10\}$}   (s3)
                    edge[directed, bend right=10]    node[above, rotate=90] {\scriptsize $b(e)=\{2,3,4\}, d(e)=\{9,10\}$}   (s4)
            (s3)    edge[directed, bend right=10]    node[above, rotate=-90] {\scriptsize $b(e)=\{2,3,4\}, d(e)=\{9,10\}$}   (s1)
                    %edge[directed, bend right=10]    node[below, rotate=45] {$b(e)=\{4,5,6\}, d(e)=\{1,2\}$}   (s2)
                    edge[directed, bend left=10]    node[above] {\scriptsize $b(e)=\{2,3,4\}, d(e)=\{9,10\}$}   (s4)
            (s4)    edge[directed, bend left=10]    node[below, rotate=-45] {\scriptsize $b(e)=\{2,3,4\}, d(e)=\{9,10\}$}   (s1)
                    edge[directed, bend right=10]    node[above, rotate=-90] {\scriptsize $b(e)=\{4,5,6\}, d(e)=\{1,2\}$}   (s2)
                    edge[directed, bend left=10]    node[below] {\scriptsize $b(e)=\{2,3,4\}, d(e)=\{9,10\}$}   (s3);
\end{tikzpicture}
    \caption{Example dynamic link network}
    \label{fig:ex_dyn_net}
\end{figure}
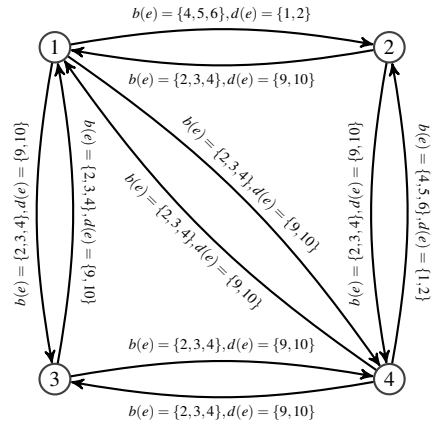

 Static networks can be modeled as (directed) weighted graphs $(V, E, p_1,\ldots,p_k)$, where $V$ is a set of nodes, $E\subseteq V\times V$ is a set of directed edges, and $p_i$ is a link parameter function $p_i: E \to \mathds{N}$, for $i\in\{1,\ldots,k\}$; without loss of generality, we assume that the parameter functions map to non-negative integers (denoted by $\mathds{N}$) or related values can be encoded with them. Similarly, dynamic link networks can be modeled as such graphs, however, $p_i$ maps an edge to a non-empty set of integer values, i.e., $p_i: E \to 2^{\mathds{N}}\setminus\emptyset$, where $2^{\mathds{N}}$ denotes the power-set of $\mathds{N}$. An example dynamic network is depicted in Fig.~\ref{fig:ex_dyn_net}, and its model $\mathcal{N}=(V,E,p_1(e),p_2(e))$, where:
 
\[
\begin{footnotesize}
\begin{aligned}
V&=\{1,2,3,4\}\\
E&=\{(1,2),(2,1),(1,3),(3,1),(1,4),\\
\hspace{0.5cm} &\hspace{0.5cm}
(4,1),(2,4),(4,2),(3,4),(4,3)\}\\
p_1(e)&=b((s,d))=\begin{cases}
        \{4,5,6\}, & \text{if } d = 2\\
        \{2,3,4\}, & \text{otherwise}
        \end{cases}\\ 
p_2(e)&=d((s,d))=\begin{cases}
        \{1,2\}, & \text{if } d = 2\\
        \{9,10\}, & \text{otherwise}
        \end{cases}.
\end{aligned}    
\end{footnotesize}
\]

Semantically, this model represents a dynamic link network in which the link's available bandwidth can vary according to the function $b$ (for \emph{bandwidth}), and the link's delay can vary according to the function $d$ (for \emph{delay}). %Note that a dynamic link network snapshot, at a given time instance, is a static network, and thus, we use both terms interchangeably.

\paragraph{Virtualization \& software technologies involved}

%As modern network systems become more complex, it is no longer enough for researchers to only use pure analytical models to carry out the evaluation and performance analysis of new services or protocols. 
Nowadays, virtualization technologies \cite{giallorenzo2021virtualization} have been used not only for the design of these networks, but also to conduct necessary experiments (emulations). The reason is that virtualization allows creating controlled environments (virtual instances) and reproduce their real underlying network architectures, as well as creating artificial conditions, that can be hard to reproduce in real life.  %aiming to test different characteristics. 
Below we present the virtualization technologies, utilized in our work, namely Virtual Machines, Containers, and  Unikernels.

\textbf{Virtual Machines (VMs)} are created and managed by a hypervisor \cite{9556458}, which allows to fully emulate different types of devices (e.g., routers, cellphones, etc.) with their own complete and isolated operating system (OS) or CPU architecture. 

\textbf{Containers} are isolated environments (user-spaces instances) for processes; they are created by utilizing the kernel features of a given operating system \cite{8693491}. In contrast to VMs, containers do not get their own virtualized hardware but use the hardware of the host system. Therefore, not having to emulate hardware and boot a complete operating system enables containers to be more efficient than VMs but dependent on the kernel functions of the host OS. 

\textbf{Unikernels} are single-address-space machine images constructed by using library operating systems \cite{8861307}. The approach consists of packaging a given application (e.g., a router) with the minimal kernel functions and drivers required to run as a sealed and immutable executable directly on the hardware (or sometimes on a hypervisor). 

\begin{figure}[!hbt]
\centering
    \begin{subfigure}[b]{0.23\textwidth}
    \centering
         \includegraphics[width=0.95\textwidth]{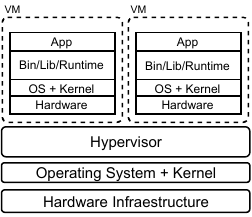}
         \caption{Virtual Machine}
         \label{subfig:VM}
    \end{subfigure}
    \begin{subfigure}[b]{0.23\textwidth}
    \centering
         \includegraphics[width=0.95\textwidth]{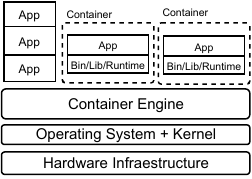}
         \caption{Container}
         \label{subfig:container}
    \end{subfigure}
    \\[1ex]
    \begin{subfigure}[b]{0.23\textwidth}
    \centering
         \includegraphics[width=0.95\textwidth]{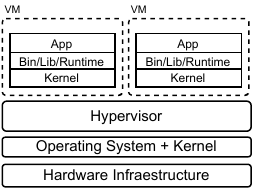}
         \caption{Unikernel}
         \label{subfig:unikernel}
    \end{subfigure}
    \\[1ex]
	\caption{Virtualization Architectures: VM (a), Container (b), Unikernel (c)}
	\label{fig:Virtualization}
\end{figure}

In Figure~\ref{fig:Virtualization}, we depict the architectural differences between the virtualization technologies. Although, the minimality of unikernels seems to be a promising emulation solution between the flexibility of virtual machines and the performance of containers, it requires applications to be written specifically for them. Thus, we do not consider this option in our work. At the same time, plenty of emulators as introduced above rely just on one single virtualization technology to emulate homogeneous networks. In contrast, we believe that both virtualization technologies (VMs and containers) can be used simultaneously (hybrid approach), to emulate heterogeneous nodes and thus replicate other types of networks with dynamic capabilities.

% Deleted , saved for final report 

% Namespaces are a feature of the Linux kernel that partitions kernel resources such that one set of processes sees one set of resources and another set of processes sees a different set of resources. The feature works by having the same namespace for a group of resources and processes, but those namespaces refer to distinct resources.

%Control groups (Cgroups) are used for scheduling and resource management so that its possible to limit and isolates the resource (CPU, memory, disk I/O) usage for all processes  belonging to a virtual machine or a container by getting its process identification number (PID) and applying an attribute in the cgroup hierarchy, Also its is used for monitoring purposes. 

\section{\uppercase{Emulator Architecture}}\label{sec:arch}

\begin{figure*}[!htb]
	\centering
	\includegraphics[width=0.7\textwidth]{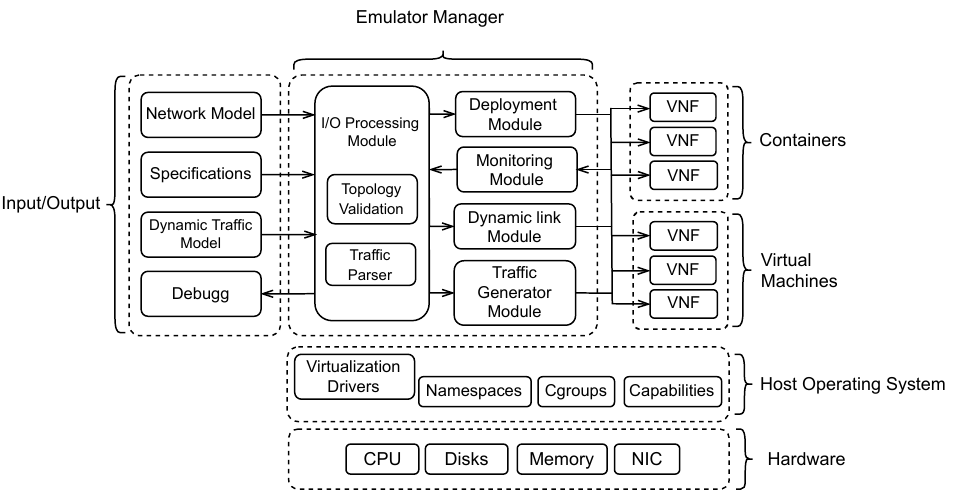}
	\caption{Emulation Platform Architecture}
	\label{fig:emulator_architecture}
\end{figure*}

Our emulation platform design and architecture is based on well-known state-of-the-art technologies presented above, such as virtualization (VMs and containers) and Linux kernel features (namespaces or cgroups). When combined efficiently, these technologies provide excellent capabilities for the emulation of a diverse set of network topologies alongside the dynamic links and interconnected network devices. The emulation platform architecture, shown in Figure~\ref{fig:emulator_architecture},  consists of several independent, flexible and configurable components. We describe each of these components in detail in the following paragraphs. 

The \textbf{Emulator Manager} is the main component and the central processing unit. It has a single instance per physical machine and it is composed of several independent modules in charge of the management, deployment and verification of the emulator components for a given network description (input for the emulation). In addition, it is responsible for providing, within the same physical host, the containers or virtual machines required for each emulated device as well as their own emulated network specifications. 

The \textbf{Input/Output Processing Module} fulfills several tasks. First, since our emulation platform relies on state-of-the-art virtualization (or container-based) solutions, it is in charge of creating and maintaining a network model that later is used by other modules to implement the necessary infrastructure elements for each emulation. To achieve this, we utilize a formal network description (specification) in terms of first order logic formulas verified throughout the emulation by a satisfiability modulo theories solver. Indeed, the network topology can be verified using model checking strategies before its actual implementation, as well as at run-time, to assure that certain properties of interest hold for the static network instances \cite{emul_verif_nca2020}. Finally, the module is also in charge of parsing and verifying the file to generate dynamic traffic scenarios between the components of an emulated network as well as the debugging output of the platform. An example of a network description is given in Listing 1 (for the network in Figure~\ref{fig:ex_dyn_net}.%, while an example of a traffic scenario is shown in Listing 2.

%\vspace{0.5em}
%\begin{adjustbox}{width=20em,height=25em,keepaspectratio}
\begin{lstlisting}[basicstyle=\scriptsize,caption={Example of a Network description (SMT-LIB)}, captionpos=b, label={lst:ex_model}]

(declare-datatypes() 
    ((Edge (mk-edge (src Int) (dst Int)))))
(declare-fun bandwidth (Edge) Int)  
(declare-fun delay (Edge) Int)
;; Node storage omitted on purpose 
;; to reduce the space, see edge storage
(declare-const edges (Array Int Edge))
(declare-const edges_size Int)
(assert (= (store edges 1 (mk-edge 1 2)) edges))
;; Edge storage omitted on purpose 
;; to reduce the space, see first and last edge
(assert (= (store edges 10 (mk-edge 4 2)) edges))
(assert (= edges_size 10))
(assert (forall ((x Int)) (=>
      (and (> x 0) (<= x edges_size))
      (and
        (=> (= (dst (select edges x)) 2) 
        ;; ite not used on purpose
            (and
                (>= (bandwidth (select edges x)) 4)
                (<= (bandwidth (select edges x)) 6)
                (>= (delay (select edges x)) 1)
                (<= (delay (select edges x)) 2)
            )
        )
        (=> (not (= (dst (select edges x)) 2))
            (and
                (>= (bandwidth (select edges x)) 2)
                (<= (bandwidth (select edges x)) 4)
                (>= (delay (select edges x)) 9)
                (<= (delay (select edges x)) 10)
            )
        ) ) ) ) ) ;; closing parentheses

\end{lstlisting}
%\end{adjustbox}
%\vspace{0.5em}

The \textbf{Deployment Module} is in charge of converting the previously generated network model into running instances of emulated network devices. In order to achieve this, the module makes use of the \verb|docker| engine for the management and support of containers and \verb|libvirt| for different virtualization technologies such as \verb|KVM|, \verb|VMware|, \verb|LXC|, and \verb|virtualbox|. At the first step, it takes the input specification and creates the required nodes with their corresponding images and properties. Each emulated node is deployed by means of a VM or a container attached to its own namespace and acts according to the software or service running inside of it (as requested by the input specification). For example, if it is desired to run a virtual switch as a Docker container, the Deployment Module creates the proper container and executes the corresponding Virtual Network Function (VNF) via a docker image (e.g., Open vSwitch). Therefore, each node has an independent view of the system resources such as process IDs, user names, file systems and network interfaces while still running on the same hardware. It can also hold several individual (virtual) network interfaces, along with its associated data, including ARP caches, routing tables and independent TCP/IP stack functions.

This gives great flexibility and capabilities to the emulator, it can execute any real software as in the real physical systems. At the last step, the module creates the links between the nodes to complete the emulation topology. The links are emulated with Linux virtual networking devices; \verb|TUN/TAP| devices are used to provide packet reception and transmission for user space processes (applications or services) running inside each node. They can be seen as simple Point-to-Point or Ethernet devices, which, instead of receiving (and transmitting, correspondingly) packets from a physical medium, read (and write, correspondingly) them from a user space process. \verb|veth| (virtual Ethernet) devices are used for combining the network facilities of the Linux kernel to connect different virtual networking components together. \verb|veth| are built as pairs of connected virtual Ethernet interfaces and can be thought of as a virtual ``patch'' cable. Thus, packets transmitted on one device in the pair are immediately received on the other device and when either device is down the link state of the pair is down too. 

%\vspace{0.6em}
%\begin{adjustbox}{width=25em,height=18.5em,keepaspectratio}
\begin{lstlisting}[basicstyle=\scriptsize,caption={Example of a Dynamic Traffic Description}, captionpos=b, label={lst:ex_traffic}]
{ "name": "test_traffic_scenario",
 "initialTime": 0,
 "timeUnitInSec": 1e-3,
 "bandwidthUnitInBits": 1e3,
    "flowSequence": [{
	  "time": 4000,
	  "flows": [{
		"flow_id": 1,
		"requiredBandwidth": 2,
		"protocol": "tcp",
		"source_ip": "192.168.1.2",
		"source_port": 4000,
		"destination_ip": "192.168.1.3",
		"destination_port": 4000
	   }]
	   },
		{
		"time": 10000,
		"flows": [{
		"flow_id": 1,
		"requiredBandwidth": 4,
		"protocol": "tcp",
		"source_ip": "192.168.1.2",
		"source_port": 8000,
		"destination_ip": "192.168.1.3",
		"destination_port": 8000
		},{
		"flow_id": 2,
		"requiredBandwidth": 3,
		"protocol": "tcp",
		"source_ip": "192.168.1.2",
		"source_port": 8080,
		"destination_ip": "192.168.1.3",
		"destination_port": 8080
		}]
	}]
}

\end{lstlisting}
%\end{adjustbox}
%\vspace{0.6em}

\begin{figure}[!htb]
	\centering
	\includegraphics[width=0.4\textwidth]{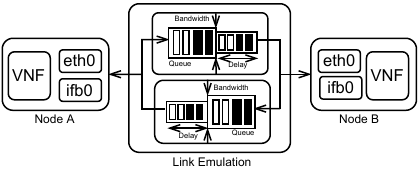}
	\caption{Asymmetric Link emulation model}
	\label{fig:link_emulation}
\end{figure}

The \textbf{Dynamic Link Module} is in charge of establishing and modifying the dynamic properties of the links (between the nodes) during the emulation's execution time. An asymmetric link between two nodes, as shown in Figure~\ref{fig:link_emulation}, is emulated by a set of nesting queues; in the simplest case - two queues. 

At the first step, packets are queued or dropped depending on the size of the first queue. This queue is drained at a rate corresponding to the link's bandwidth. Once outside, packets are staged in a delay line for a specific time (propagation delay of the link) in the second queue and then finally injected into the network stack. This module uses the Linux Advanced traffic control \verb|tc|, to control and set these properties by using filtering rules (classes) to map data (at the data link or the network layer) to queuing disciplines (\verb|qdisc|) in an egress network interface. Note that since \verb|tc| can be used only on egress, traffic Intermediate Functional Block devices (IFB) are created to allow queuing disciplines on the incoming traffic and thus use the same technique.

The \textbf {Traffic Generation Module} is in charge of converting the dynamic description of the traffic (see an example of it in Listing 2), into a timed sequence of network packets. This sequence is then introduced into the deployed nodes during the emulation. For the generation of network packets, the module uses \verb|nmap| at each node, particularly \verb|nping|, allowing to generate traffic with headers from different protocols. This is achieved by using \verb|virsh| commands using \verb|libvirt| for virtual machines or by passing \verb|execute| commands through the Docker daemon (for containers). It is important to keep in mind that \verb|nping| can be replaced for any other software to generate traffic. Additionally, multiple instances of the same or different traffic generators can be executed inside each emulated node. 

Finally, the \textbf{Monitoring Module} retrieves and collects information from the nodes and their links. This information can be used for example, to verify that the emulation process is executed correctly. Additionally, for our case study, this information is useful to change the bandwidth on demand (in fact this is known as demand assigned multiple access or \emph{DAMA} which further serves as a case study).

\section{\uppercase{Evaluation}}\label{sec:eval}
%In order to showcase the dynamic link capabilities of our emulator, we present an experimental evaluation, varying different network parameters i.e., bandwidth and delay. We present this experimental evaluation through a real use case which requires the flexibility as provided by our emulator.

\paragraph{Use case -- Satellite Communications}
It is certainly difficult to assess novel software systems in satellite networks,  mainly due to the complex network dynamics, models and mechanisms that are involved in such networks. Particularly, the difficulties are related to the asymmetric bandwidth capacities of satellite links (due to the use of different radio-frequencies and modulations) and the highly dynamic behaviour of the mechanisms to make an efficient use of the scarce and costly transmission resources. Usually, these mechanisms tend to dynamically allocate bandwidth depending on internal factors but, more importantly external factors (e.g., traffic crossing the network) can also be crucial.

We focus on the DAMA mechanism for dynamic link bandwidth assignation.
As its name suggests, the idea is to assign physical resources to nodes which need to transmit traffic, according to their demands. The physical resources are translated to bandwidth. In real settings, in order to assign a given bandwidth, the DAMA module may assign certain time slots and frequencies to a particular node (station). We abstract from this physical level and consider that the bandwidth to physical resource assignation is done by a proper translation module. Therefore, we are interested in the logical view of the algorithm, i.e., only the bandwidth assignations and no deeper. It is important to note that in DAMA, nodes have minimal and maximal bandwidth capacities assigned. Likewise, the satellite transponder has a maximal bandwidth capacity that it can handle. %For more details, an appendix is left for the reviewers on Appendix~\ref{sec:dama}.
%utilize different types of traffic assignment techniques that may be combined in order to both ensure a high utilisation of the return link resources and to offer different  capacity assignment types.
We assume that the behaviour of user traffic is not predictable in advance, i.e., we do not deal with a seasonal behavior of the network traffic. Thus, dynamic bandwidth assignations of the emulation system are required. That is, the passing traffic modifies the configuration of the emulation and consequently the resulting modifications will have an effect on the traffic going through the emulator. The goal of the developed emulator is to reproduce the conditions described above.

\paragraph{Experimental Settings \& Results}

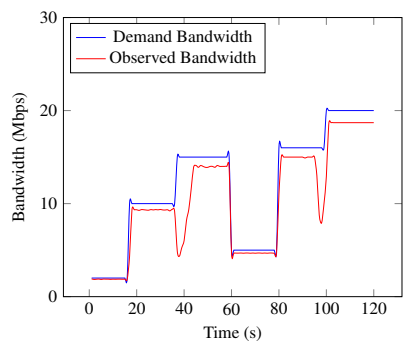
\begin{figure}[!htb]
    \centering
    \begin{tikzpicture}[scale=0.65]
\begin{axis}[
    ymax=30, 
    ymin=0,
	ylabel={Bandwidth (Mbps)},
	xlabel={Time (s)},
	legend style={at={(0.02,0.98)}, 
	anchor=north west},
]
\addplot[smooth,blue] coordinates
{

(1.00 , 2)
(2.00 , 2)
(3.00 , 2)
(4.00 , 2)
(5.00 , 2)
(6.00 , 2)
(7.00 , 2)
(8.00 , 2)
(9.00 , 2)
(10.00 , 2)
(11.00 , 2)
(12.00 , 2)
(13.00 , 2)
(14.00 , 2)
(15.00 , 2)
(16.00 , 2)
(17.00 , 10)
(18.00 , 10)
(19.00 , 10)
(20.00 , 10)
(21.00 , 10)
(22.00 , 10)
(23.00 , 10)
(24.00 , 10)
(25.00 , 10)
(26.00 , 10)
(27.00 , 10)
(28.00 , 10)
(29.00 , 10)
(30.00 , 10)
(31.00 , 10)
(32.00 , 10)
(33.00 , 10)
(34.00 , 10)
(35.00 , 10)
(36.00 , 10)
(37.12 , 15)
(38.02 , 15)
(39.00 , 15)
(40.01 , 15)
(41.00 , 15)
(42.00 , 15)
(43.00 , 15)
(44.00 , 15)
(45.00 , 15)
(46.00 , 15)
(47.00 , 15)
(48.00 , 15)
(49.00 , 15)
(50.00 , 15)
(51.00 , 15)
(52.00 , 15)
(53.00 , 15)
(54.00 , 15)
(55.00 , 15)
(56.00 , 15)
(57.00 , 15)
(58.00 , 15)
(59.00 , 15)
(60.00 , 5)
(61.00 , 5)
(62.00 , 5)
(63.00 , 5)
(64.00 , 5)
(65.00 , 5)
(66.00 , 5)
(67.00 , 5)
(68.00 , 5)
(69.00 , 5)
(70.00 , 5)
(71.00 , 5)
(72.00 , 5)
(73.00 , 5)
(74.00 , 5)
(75.00 , 5)
(76.00 , 5)
(77.00 , 5)
(78.00 , 5)
(79.00 , 5)
(80.00 , 16)
(81.00 , 16)
(82.00 , 16)
(83.00 , 16)
(84.00 , 16)
(85.00 , 16)
(86.00 , 16)
(87.00 , 16)
(88.00 , 16)
(89.00 , 16)
(90.00 , 16)
(91.00 , 16)
(92.00 , 16)
(93.00 , 16)
(94.00 , 16)
(95.00 , 16)
(96.00 , 16)
(97.00 , 16)
(98.00 , 16)
(99.00 , 16)
(100.00 , 20)
(101.00 , 20)
(102.00 , 20)
(103.00 , 20)
(104.00 , 20)
(105.00 , 20)
(106.00 , 20)
(107.00 , 20)
(108.00 , 20)
(109.00 , 20)
(110.00 , 20)
(111.00 , 20)
(112.00 , 20)
(113.00 , 20)
(114.00 , 20)
(115.00 , 20)
(116.00 , 20)
(117.00 , 20)
(118.00 , 20)
(119.00 , 20)
(120.00 , 20)

};\addlegendentry{Demand Bandwidth}

%\addplot[smooth,mark=*,red] coordinates
\addplot[smooth,red] coordinates
{

(1.00 , 1.91)
(2.00 , 1.85)
(3.00 , 1.87)
(4.00 , 1.89)
(5.00 , 1.85)
(6.00 , 1.89)
(7.00 , 1.88)
(8.00 , 1.88)
(9.00 , 1.86)
(10.00 , 1.88)
(11.00 , 1.87)
(12.00 , 1.88)
(13.00 , 1.88)
(14.00 , 1.87)
(15.00 , 1.88)
(16.00 , 1.87)
(17.00 , 4.32)
(18.00 , 9.32)
(19.00 , 9.34)
(20.00 , 9.34)
(21.00 , 9.27)
(22.00 , 9.36)
(23.00 , 9.34)
(24.00 , 9.31)
(25.00 , 9.29)
(26.00 , 9.36)
(27.00 , 9.32)
(28.00 , 9.36)
(29.00 , 9.33)
(30.00 , 9.37)
(31.00 , 9.32)
(32.00 , 9.31)
(33.00 , 9.34)
(34.00 , 9.22)
(35.00 , 9.38)
(36.00 , 9.16)
(37.12 , 4.89)
(38.02 , 4.35)
(39.00 , 5.25)
(40.01 , 6.03)
(41.00 , 8.19)
(42.00 , 9.41)
(43.00 , 11.6)
(44.00 , 13.9)
(45.00 , 14.1)
(46.00 , 13.9)
(47.00 , 14.1)
(48.00 , 14.0)
(49.00 , 13.9)
(50.00 , 13.9)
(51.00 , 14.0)
(52.00 , 14.0)
(53.00 , 13.9)
(54.00 , 14.1)
(55.00 , 14.0)
(56.00 , 14.0)
(57.00 , 14.0)
(58.00 , 14.0)
(59.00 , 13.8)
(60.00 , 4.67)
(61.00 , 4.69)
(62.00 , 4.68)
(63.00 , 4.68)
(64.00 , 4.69)
(65.00 , 4.67)
(66.00 , 4.69)
(67.00 , 4.69)
(68.00 , 4.67)
(69.00 , 4.69)
(70.00 , 4.67)
(71.00 , 4.69)
(72.00 , 4.69)
(73.00 , 4.67)
(74.00 , 4.69)
(75.00 , 4.67)
(76.00 , 4.69)
(77.00 , 4.69)
(78.00 , 4.67)
(79.00 , 4.69)
(80.00 , 11.0)
(81.00 , 15.0)
(82.00 , 15.0)
(83.00 , 15.0)
(84.00 , 15.0)
(85.00 , 15.0)
(86.00 , 15.0)
(87.00 , 15.0)
(88.00 , 15.0)
(89.00 , 15.0)
(90.00 , 15.0)
(91.00 , 14.9)
(92.00 , 15.0)
(93.00 , 15.0)
(94.00 , 15.0)
(95.00 , 15.0)
(96.00 , 13.0)
(97.00 , 8.79)
(98.00 , 7.93)
(99.00 , 9.98)
(100.00 , 12.9)
(101.00 , 18.5)
(102.00 , 18.7)
(103.00 , 18.7)
(104.00 , 18.7)
(105.00 , 18.7)
(106.00 , 18.7)
(107.00 , 18.7)
(108.00 , 18.7)
(109.00 , 18.7)
(110.00 , 18.7)
(111.00 , 18.7)
(112.00 , 18.7)
(113.00 , 18.7)
(114.00 , 18.7)
(115.00 , 18.7)
(116.00 , 18.7)
(117.00 , 18.7)
(118.00 , 18.7)
(119.00 , 18.7)
(120.00 , 18.7)

};\addlegendentry{Observed Bandwidth}
\end{axis}
\end{tikzpicture}
    \caption{Varying emulated bandwidth by demand}
    \label{fig:band}
\end{figure}

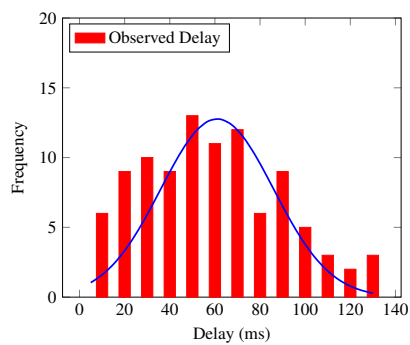
\begin{figure}[!htb]
    \centering
    \begin{tikzpicture}[scale=0.65]
\begin{axis}[
             ymin=0,
             ymax=20,
             xlabel={Delay (ms)},
	         ylabel={ Frequency }, % before estimated probability density function (PDF) f(delay) 
	         legend style={at={(0.02,0.98)}, anchor=north west}]

\addplot[ybar, bar width=5, red, fill, area style] coordinates { 
(10,	    6)
(20,	9)  
(30,	10)
(40,	9)
(50,	13)
(60,	11)
(70,	12)
(80,	6)
(90,	9)
(100,	5)
(110,	3)
(120,	2)
(130,	3)
};\addlegendentry{Observed Delay}

\addplot[yscale=800,smooth, no marks,line width=1pt, blue,domain={5:130}]{\gauss{61}{25}}; %\addlegendentry{Delay}
% took it from this post https://tex.stackexchange.com/questions/237085/histogram-with-overlaying-gauss-curve-pgfplots 

\end{axis}
\end{tikzpicture}
    \caption{Emulated delay histogram}
    \label{fig:delay}
\end{figure}

For the link's bandwidth assignation we configured it by demand, that is the link's bandwidth gets assigned to whatever the node requires up to a maximal allocation (as in satellite communications). The results are presented in Figure~\ref{fig:band}; as can be seen, the measured bandwidth (with the \verb|bmw-ng| utility) varies considerably w.r.t. the demand.
%With the bandwidth by demand, the assigned bandwidth cannot exceed an established constant known as MIR (maximum information rate) and it is guaranteed to be assigned at least a CIR (committed information rate); additionally, for all the nodes in the cluster, a total bandwidth capacity (denoted as $B$) is set to 20 Mbps. In our experiments the CIR was set to 2 Mbps MIR to 4 Mbps for each node in the cluster of a 4 node mesh, and $B$ was set to 20 Mbps. The demand was set between 1 to 16 Mbps. 
%\subsection{Delay} 
%In order to showcase the flexibility of 
For the link's delay emulation, we have varied the desired delay between 5ms and 100ms. The assignation was made randomly by setting the delay queue with a \emph{normal} distribution (using the \verb|tc| utility). The results are presented in Figure~\ref{fig:delay}; as can be seen, the delays have been measured (with \verb|nping|)
and the histogram clearly shows a normal distribution, as expected. It is important to note that the measured delay can be higher than 100ms since the processing time of the equipment plays a role in the delay, as in the real system.

In this section, we have showcased how both link parameters of interest (bandwidth and delay) can be dynamically (and automatically) assigned in our emulator; it can properly emulate a complex DAMA scheme for satellite communications. Moreover, it is capable of producing distinct traffic configurations to thoroughly test novel engineered solutions. Additionally, the emulator's design allows to easily interact with real life software components.

% Also idea .  I think we can move the last paragrah of section 5.1 at the begining of section 5.2 or maybe here ... not sure.

\section{\uppercase{Conclusion}}\label{sec:conc}
In this paper, we have showcased the design and architecture for a dynamic link network emulator, which can be used for creating complex networks, such as for example satellite communication networks. The emulator is flexible and can emulate any existing software; additionally, it can dynamically change the link parameter values. At the moment, mobile nodes were not considered which means networks whose topology may change (such as MANETS) are not supported by the emulator.

As for future work, we plan to incorporate more features into the architecture so that it becomes more controllable and realistic, for example, we consider incorporating link state scenarios to qualify the solutions under different conditions (degraded links, weather conditions, mobility etc.). Additionally, we plan to investigate various model-based simulation strategies for dynamic link networks, to efficiently (faster than the emulation) obtain static network snapshots with certain desired properties; this should allow the emulator to provide interesting network states to assess novel solutions.

% --------- References -------------------

\bibliographystyle{apalike}
{\small
\bibliography{references}}
%\bibliography{references}

%\appendix\clearpage
%\section{Demand Assigned Multiple Access details}\label{sec:dama}
%\input{sections/dama}

\end{document}